\documentclass[prl,twocolumn,amsmath,amssymb,floatfix]{revtex4}
\usepackage{graphicx}
\usepackage{dcolumn}
\usepackage{bm}

\newcommand{\lvo}{$\rm{LiV_2O_4}$}
\newcommand{\avo}{$\rm{AV_2O_4}$}
\newcommand{\zvo}{$\rm{ZnV_2O_4}$}
\newcommand{\zco}{$\rm{ZnCr_2O_4}$}

\newcommand{\dxy}{$d_{xy}$}
\newcommand{\dyz}{$d_{yz}$}
\newcommand{\dzx}{$d_{zx}$}
\newcommand{\neel}{$\rm N\acute{e}el$}

\begin{document}
\preprint{BAPS/123-QED}

\title{Orbital and spin chains in $\rm {\bf ZnV_2O_4}$}

\author{S.-H. Lee$^1$}
\author{D. Louca$^2$}
\author{H. Ueda$^3$}
\author{S. Park$^4$}
\author{T.J. Sato$^3$}
\author{M. Isobe$^3$}
\author{Y. Ueda$^3$}
\author{S. Rosenkranz$^5$}
\author{P. Zschack$^6$}
\author{J. \'I\~niguez$^1$}
\author{Y. Qiu$^1$}
\author{R. Osborn$^5$}

\affiliation{%
$^1$NIST Center for Neutron Research, National Institute of
Standards and Technology, Gaithersburg, MD 20899\\
$^2$Department of Physics, University of Virginia, Charlottesville, VA 22904\\
$^3$Institute for Solid State Physics, University of Tokyo, Kashiwa, Chiba 277-8581, Japan\\
$^4$HANARO Center, Korea Atomic Energy Research Institute, Daejeon, Korea\\
$^5$Material Science Division, Argonne National Laboratory,
Argonne, IL 60439\\
$^6$Frederick-Seitz Materials Research Lab, University of Illinois
at Urbana-Champaign, IL 61801}

\date{\today}

\begin{abstract}
Our powder inelastic neutron scattering data indicate that \zvo~is
a system of spin chains that are three dimensionally tangled in
the cubic phase above 50 K due to randomly occupied $t_{2g}$
orbitals of V$^{3+}$ ($3d^2$) ions.  Below 50 K in the tetragonal
phase, the chains become straight due to antiferro-orbital
ordering. This is evidenced by the characteristic wave vector
dependence of the magnetic structure factor that changes from
symmetric to asymmetric at the cubic-to-tetragonal transition.

\end{abstract}


\maketitle

Cubic spinels, AB$_2$O$_4$, with magnetic B ions have attracted
considerable attention recently in light of geometrical
frustration intrinsic to the B-site sublattice of corner-sharing
tetrahedra \cite{ramirez2001}. The long sought zero energy mode of
spin fluctuations in the B sublattice
\cite{moessner1998,canals1998} was found in \zco \cite{shl2002}.
Also in \zco, a spin-Peierls-like transition to relieve
frustration was found upon cooling where a tetragonal distortion
and a magnetic long range order occur simultaneously
\cite{shl2000}. When the B site is occupied by vanadium ions with
orbital degeneracy, complex electronic and magnetic properties
emerge. \lvo, for instance, with monovalent Li ions at the
tetrahedral A site and mixed valent V$^{3.5+}$ ions exhibits heavy
fermion (HF) behavior at low temperatures with the largest
Sommerfeld constant among $d$-electron systems, $\gamma \approx
0.42$ J/mol K$^2$\cite{kondo}. \avo~ with divalent ions such as Zn
\cite{ueda1997}, Mg \cite{mamiya1997}, Cd \cite{onoda2002}, at the
A site and trivalent V$^{3+}$ ($3d^2$) ions is a Mott insulator
\cite{fujimori1988} that undergoes two separate phase transitions
at low temperatures, in contrast to other insulating spinels
without orbital degeneracy such as \zco. In this Letter, we show
that the orbital degree of freedom plays the central role in the
physics of vanadates.

Many theoretical efforts have been made to understand the unusual
low temperature behaviors of metallic and insulating vanadates.
The macroscopic ground state degeneracy induced by the geometrical
frustration intrinsic to the magnetic lattice was attributed to
explain the enhancement of the specific heat at low temperatures
in \lvo~\cite{eyert1999}. It was also used to explain why the
\neel~temperature, $T_N$, is considerably lower than the
Curie-Weiss temperature, $\Theta_{CW}$, in the insulating
vanadates. Spin-lattice coupling mechanisms have been proposed to
explain the phase transitions of
\zvo~\cite{yamashita2000,tchernyshyov2002}, but fail to explain
why the spin and lattice order at different temperatures in the
insulating vanadates unlike in \zco. Orbital degeneracy of the
vanadium ions was recently considered as well. Fulde {\it et al.}
proposed \cite{fulde2001} that due to frustrated charge order or
orbital order in \lvo~$(3d^{1.5})$, one-dimensional chains form,
that contribute to the enhancement of the linear term in the
specific heat. For insulating \avo~(A = Zn, Mg, Cd) $(3d^2)$,
Tsunetsugu and Motome proposed \cite{tsunetsugu2003} that in the
tetragonal ($c < a = b$) phase, among the triply degenerate
$t_{2g}$ orbitals, the $d_{xy}$ orbital is favored and is occupied
by one electron at every V site. The second electron is in an
antiferro-orbital state that can be described by stacking the
$ab$-planes along the $c$-axis with alternating $d_{yz}$ and
$d_{zx}$ orbitals. This effectively forms straight spin chains on
the $ab$-planes. Using a crystal symmetry argument, a
ferro-orbital model for the orbital state of the second electron
was also proposed \cite{khomskii,tchernyshyov2004}. To test the
validity of these theoretical models, detailed studies of magnetic
correlations in the vanadates are necessary to elucidate the
interplay between spin, orbital, and lattice degrees of freedom.

The insulating \zvo~($S=1$) exhibits a sharp drop in the bulk
susceptibility, $\chi$, at 50 K \cite{ueda1997}. A magnetic long
range order occurs at 40 K with a characteristic wavevector of
$\vec{Q}=(110)$ and an ordered moment $\langle M\rangle = g\langle
S\rangle $ = 0.61(3) $\mu_{B}/V$ (Fig. 1), in spite of strong
magnetic interactions evidenced by the large $|\Theta_{CW}| =
JzS(S+1)/3k_{B}$ = 998(5) K \cite{niziol1973,reehuis2003}. The
sharp drop in $\chi$ is associated with a structural transition
from a high temperature cubic ($a=8.39941(5)$ \AA) to a low
temperature tetragonal phase ($a_{tet}=5.94807(5)$ \AA~$\approx
a/\sqrt{2}$ and $c_{tet}=8.37532(1)$ \AA) (see Fig. 1 (b)). From
our neutron scattering measurements, we find that when the
tetragonal distortion occurs, the wave vector ($Q$) dependence of
the inelastic magnetic neutron scattering of the powder sample
changes lineshape from symmetric centered at $Q_c^{cub}=$ 1.35(4)
\AA$^{-1}$ to asymmetric peaked at $Q_c^{tet}=$ 1.10(2)
\AA$^{-1}$. Quantitative analysis shows that \zvo~is a system of
three-dimensionally tangled spin chains in the cubic phase. On the
other hand, in the tetragonal phase \zvo~becomes an excellent
model system for one-dimensional spin chains. This favors the
antiferro-orbital model that yields straight chains in the
$ab$-planes with {\it weak} interchain interactions. We argue that
our findings provide a unified picture of the physics of
vanadates, both insulating and metallic.

A 30 g polycrystalline sample of \zvo~was used for the neutron
scattering experiments. The elastic measurements were performed
using the cold neutron triple-axis spectrometer SPINS at the
National Institute of Standards and Technology Center for Neutron
Research with a fixed incident and scattered neutron energy of
$E_{i}$ = 3.1 meV. The inelastic measurements were carried out on
the time-of-flight spectrometer LRMECS with a detector arrangement
covering scattering angles from 7.5$^o$ to 118$^o$, at the Intense
Pulsed Neutron Source of Argonne National Laboratory and with an
incident energy, $E_{i}$ = 30 meV.

\begin{figure}
\centering
\includegraphics[width=7cm]{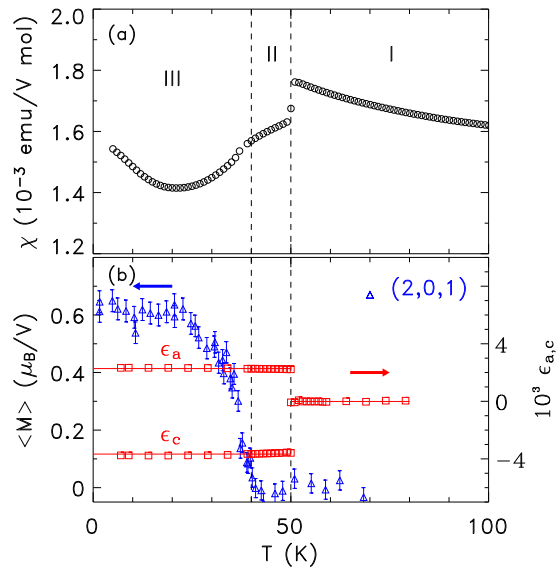}
\caption{Temperature dependence of bulk susceptibility, elastic
magnetic intensity and lattice strain. (a) Zero field cooled bulk
susceptibility, $\chi$. (b) Elastic magnetic intensity (triangles)
at $Q =$ 1.67 \AA$^{-1}$ that corresponds to the magnetic (2,0,1)
reflection with a saturation moment of $\langle M\rangle =
g\langle S\rangle $ = 0.61(3) $\mu_{B}$ per V ion
($g$-gyromagnetic ratio), and lattice strain (squares) along $a$
and $c$ measured by synchrotron X-ray diffraction on a single
crystal with dimensions of $10^{-3}$ mm$^3$ (6 $\mu$g). The X-ray
measurements were carried out at the 33BM-C beamline at the
Advanced Photon Source of Argonne National Laboratory.  }
\end{figure}



\begin{figure}
\centering
\includegraphics[width=7cm]{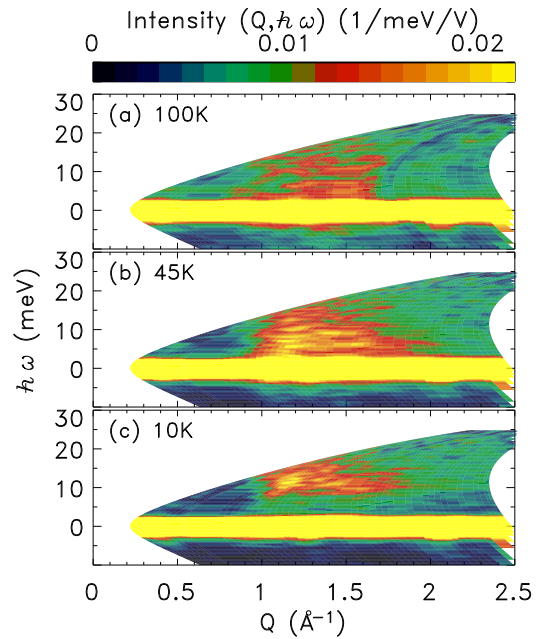} 
\caption{Neutron scattering intensity as a function of energy
($\hbar\omega$) and wave vector ($Q$) transfer obtained from a
powder sample of \zvo~at three different phases. (a) At 100 K, in
the cubic and cooperative paramagnetic phase.  (b) At 45 K, in the
tetragonal phase without magnetic long range order. (c) At 10 K,
in the tetragonal and N$\rm \acute{e}$el phase. }
\end{figure}
Fig. 2 is an overview of the inelastic neutron scattering data in
the form of color images of intensity, $I (Q,\hbar\omega)$, as a
function of $Q$ and energy transfer, $\hbar\omega$, in three
different phases. Data were collected up to $Q=6.5$ \AA$^{-1}$ but
shown only up to 2.5 \AA$^{-1}$ in Fig. 2. In the cubic phase I
($T>50$ K), strong low energy magnetic excitations are present in
the form of a broad peak centered at $Q_c^{cub}$ = 1.35(4)
\AA$^{-1}$ shown in Fig. 2(a). In phase II (40 K $< T <$ 50 K)
with the tetragonal distortion but no magnetic long-range order, a
similar broad peak is present at low energies. However, the broad
peak is strikingly asymmetric in $Q$ and shifts to a lower
characteristic wavevector, $Q_c^{tet} = 1.10(2)$ \AA$^{-1}$ (Fig.
2 (b)). In the tetragonal N$\rm \acute{e}$el phase III (T $<$ 40
K), the asymmetry of the broad feature in $Q$ remains but spectral
weight in the inelastic scattering cross section shifts in energy
to have a broad feature peaked at around 11 meV (Fig. 2 (c)). The
change from symmetric to asymmetric $Q$-dependence of the spin
excitations between 100 and 45 K indicates that there is a
crossover in the nature of the magnetic correlations from three
dimensions to a lower dimension \cite{warren1941} between phases I
and II.

\begin{figure}
\centering
\includegraphics[width=7cm]{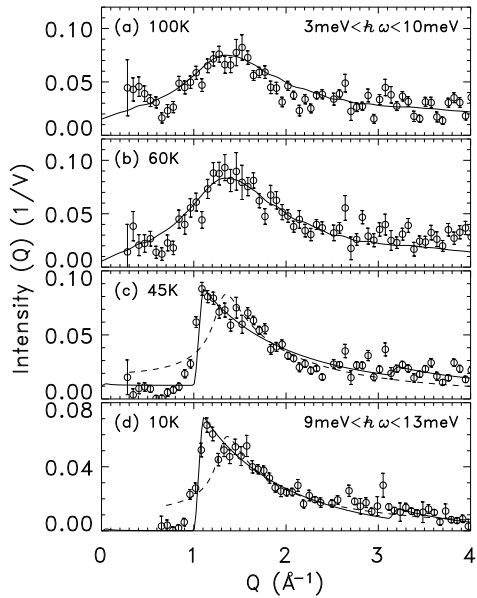}
\caption{$Q$-dependence of inelastic magnetic neutron scattering
intensity obtained by integrating the data shown in Fig. 1 over
the range of energy that includes most of the excited spectral
weight: (a) at 100 K, (b) 60 K, (c) 45 K, and (d) 10 K. The energy
integration range is 3 meV $<$ $\hbar\omega$ $<$ 10 meV for (a),
(b) and (c) and 9 meV $<$ $\hbar\omega$ $<$ 13 meV for (d). Lines
are fits to the models explained in the text.}
\end{figure}
The energy integrated inelastic magnetic neutron scattering
intensity as a function of $Q$ is shown at several temperatures in
Fig. 3. The phonon contribution was determined first at 100 K by
fitting the data to a $Q^2$ term for $Q > 5$ \AA$^{-1}$, was
multiplied by the thermal population factor to estimate the
contribution at lower temperatures, and was subtracted from the
data. In phase I of the 60 and 100 K data, the broad peak centered
around 1.35(4) \AA$^{-1}$ can be attributed to cooperative
paramagnetic spin fluctuations induced by geometrical frustration
intrinsic to the magnetic lattice. Since magnetism in both
\zco~$(3d^3)$ and \zvo~$(3d^2)$ involves $t_{2g}$ electrons, one
may think that their magnetic fluctuations should have the same
fundamental spin degrees of freedom. If that is the case,
antiferromagnetic hexagonal spin loops are the fundamental spin
degrees of freedom and will produce magnetic neutron scattering
with the characteristic wave vector of $Q_c^{hex}$ = 1.5
\AA$^{-1}$ \cite{shl2002,shl2000,tchernyshyov2002}. The
$Q_c^{hex}$, however, is inconsistent with the observed peak
position ($Q_c^{cub}=1.35(4)$ \AA$^{-1}$), which tells that
dynamic spin correlations in \zvo~ are different in nature than
those in \zco. In order to explain our data, we considered a model
similar to the one proposed in Refs. \cite{fulde2001,fujimoto2002}
that takes the orbital degeneracy of $V^{3+} (3d^2)$ ions into
account. Since in the cubic phase the three $t_{2g}$ orbitals,
$d_{xy}, d_{yz}$, and $d_{zx}$, are equivalent, we assume that at
each V site their occupancy fluctuates with time with an equal
probability of 1/3. At an instant time, two out of the three
orbitals are randomly occupied at all V$^{3+}$ sites. We
considered a model with 12 $\times$ 12 $\times$ 12 cubic unit cell
with such randomly occupied $t_{2g}$ orbitals. Fig. 4 (a) shows a
schematic representation of one such unit cell. We subsequently
consider all possible magnetic interactions due to direct overlap
of the orbitals to obtain the effective fluctuating spin objects.
Shown as sky blue rods in Fig. 4 (a), the resulting fluctuating
spin objects form three-dimensionally tangled antiferromagnetic
spin chains. The solid lines in Fig. 3 (a) and (b) are the
powder-averaged resulting structure factor squared from this model
that reproduces the data well including the characteristic wave
vector $Q_c^{cub}$.

\begin{figure}
\centering
\includegraphics[width=8.5cm]{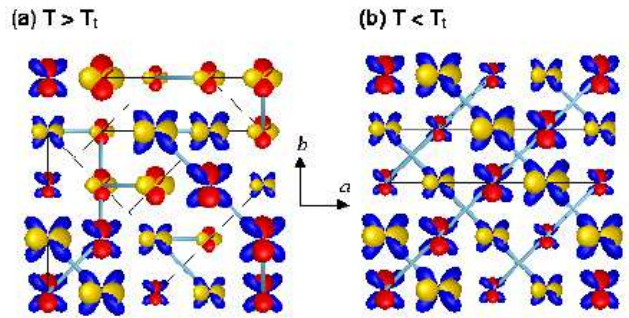} \caption{(a)-(b) Illustrations
of the orbital states of \zvo~in one cubic unit cell. Balloons
represent the $t_{2g}$ orbitals of the V$^{3+} (3d^2)$ ions:
\dxy~(blue), \dyz~(red), and \dzx~(yellow) orbitals located at the
vanadium site. The four different sizes of the ballons represent
four different $ab$-planes with different z-coordinates. (a) The
cubic phase above 50 K. The three orbitals are randomly
distributed. The blue rods connect possible dynamic magnetic
interactions at a snap shot due to direct overlap of the
neighboring orbitals. (b) Antiferro-orbital model for the
tetragonal phase. $J$, $J^{'}$, and $J_3$ represent coupling
constants for the nearest neighbor intrachain, the nearest
neighbor interchain (interplane), and the second nearest neighbor
intrachain interactions, respectively. In this model, $J{'}$ is
negligible because $d_{yz}$ and $d_{zx}$ orbitals do not overlap.}
\end{figure}
In the tetragonal phases II and III, spin fluctuations change
dramatically, with a sharp increase at low $Q$ and a long tail at
high $Q$.  It is clear that the asymmetric lineshape cannot be
explained by the two models discussed above for the cubic phase.
Instead, as shown by the solid lines in Fig. 3(c) and 3(d), the
asymmetric lineshape can be directly fit to the dynamic structure
factor based on the single mode approximation \cite{1Dchain},
\begin{equation}
S(\vec{Q},\omega) \propto |F(Q)|^2 \frac{ 1-{\rm cos}(\vec{Q}\cdot
\vec{d})}{\hbar\omega_{\vec{Q}}}\delta (\omega -
\omega_{\vec{Q}}),
\end{equation}
where $\vec{d}$ is the intrachain spacing for a one dimensional
(1D) spin chain. Here $F(Q)$ is the magnetic form factor for the
V$^{3+}$ and the spin wave dispersion relation is
$\hbar\omega_{\vec{Q}}=\sqrt{\Delta^2+v^2{\rm sin}^2(\vec{Q}\cdot
\vec{d})}$ where $\Delta$ is a spin gap and $v$ is the spin wave
velocity. The optimal intrachain spacing, $d$, obtained from the
fitting is consistent with the distance between nearest
neighboring V$^{3+}$ ions: $d= \pi/Q_c^{tet}$ = 2.97 \AA~ $
=a_{tet}/2$. The excellent agreement between the 1D chain model
and data indicates that below the tetragonal transition, \zvo~is a
system of one dimensional antiferromagnetic spin chains. This
one-dimensionality of the magnetic interactions can be understood
if two orbitals per V ion are occupied in a striated form along
the $c$-axis, as shown in Fig. 4(b). One electron from every V ion
resides in the \dxy~orbital while the occupancy of the second
electron can be described by stacking the $ab$-planes with
alternating \dyz~and \dzx~orbitals along the $c$-axis
\cite{tsunetsugu2003}. The direct overlap of neighboring $t_{2g}$
orbitals occurs only between \dxy~orbitals, yielding orbital
chains and thereby one dimensional antiferromagnetic spin chains
in the $ab$-planes. The V$^{3+}$ magnetic moments do not order
even in the orbitally ordered state because of the
one-dimensionality of the magnetic interactions until weak further
nearest neighbor interactions set in \cite{tsunetsugu2004}.
Ferro-orbital ordering where the second electron of every V$^{3+}$
ion resides on the $\frac{d_{yz}\pm {\rm i} d_{zx}}{\sqrt{2}}$
orbitals, has also been proposed \cite{tchernyshyov2004}. However,
in this model the orbitals of the neighboring second electrons do
overlap directly with each other leading to stronger interchain
interactions by at least $|J'| \sim 0.25J$ \cite{khomskii} and
this would generate a much less asymmetric lineshape in $S(Q)$
than observed \cite{ferrosq}.

Until now, \zvo~ was considered to be a geometrically frustrated
magnet similarly to \zco~ because the magnetic V$^{3+}$ ions form
a lattice of corner-sharing tetrahedra in the crystal structure.
However, our results show that due to the orbital degree of
freedom \zvo~ should be considered as a system of spin chains,
instead. In the cubic phase, the random occupancy of the $t_{2g}$
orbitals renders a system of three-dimensionally tangled spin
chains. In the tetragonal phase $(c < a = b)$, on the other hand,
antiferro-orbital ordering occurs that yields a system of weakly
interacting straight spin chains in the $ab$ planes. This picture
can also be used to explain the phase transitions observed in
other insulating vanadates, AV$_2$O$_4$ (A = Mg and Cd).  The
tetragonal transition occurs at $T_t=$ 65 K and 97 K whereas the
\neel~ ordering occurs at $T_N=$ 42 K and 35 K for Mg and Cd,
respectively. $T_t$ is determined by the balance among thermal
energy, the energy gain from orbital ordering and the energy cost
for the lattice distortion. When the A site is occupied by a
larger ion, e.g. Cd$^{2+}$ instead of Zn$^{2+}$ or Mg$^{2+}$, the
lattice becomes softer and the tetragonal distortion occurs at a
higher temperature. $T_N$, on the other hand, is determined by the
strength of interchain coupling, and therefore it is lower for
Cd$^{2+}$ where the interchain coupling is weaker due to the
larger distance between V ions.

These results may have significant implications on the physics of
metallic \lvo~ as well.
Our
finding that the cubic phase of \zvo~ consists of
three-dimensionally tangled fluctuating spin chains suggests that
the one-dimensionality of the magnetic fluctuations may play an
important role in \lvo~ that remains cubic down to 20 mK
\cite{koda2004}.  It was previously found that \lvo~ exhibits
strong antiferromagnetic spin fluctuations when the system enters
the heavy fermionic phase at low temperatures \cite{shl2001}. If
we consider that the orbital degree of freedom is important in
this system as in the cubic phase of \zvo, then the formation of
three dimensionally tangled fluctuating orbital chains may also
occur in \lvo.  The metallic character of \lvo~may produce a
spin-density-wave along the orbital chains that is responsible for
the strong antiferromagnetic spin fluctuations and the enhancement
of the low energy density of states at low temperatures.


We thank D. I. Khomskii, C. Broholm and J. B. Goodenough for
valuable discussions. This work was partially supported by the NSF
under Agreement No. DMR-9986442 and the DOE under Contracts
DE-FG02-01ER45927, W-31-109-ENG-38, DE-FG02-91ER45439, and
DE-AC05-00OR22725.

\end{document}